\documentclass[%
 reprint,
 amsmath,amssymb,
 aps,
]{revtex4-2}

\usepackage{graphicx}
\usepackage{dcolumn}
\usepackage{bm}

\usepackage{color}
\usepackage{pgfplots,mathtools}
\usepackage{hyperref}

\usepackage{braket}
\usepackage{slashed}
\newcommand{\be}{\begin{equation}}
\newcommand{\ee}{\end{equation}}
\newcommand{\bea}{\begin{eqnarray}}
\newcommand{\eea}{\end{eqnarray}}
\newcommand{\ba}[1]{\begin{array}{#1}}
\newcommand{\ea}{\end{array}}
\newcommand{\nn}{\nonumber}

\newcommand{\ep}{\epsilon}

\newcommand{\orcid}[1]{\href{https://orcid.org/#1}{\includegraphics[width=8pt]{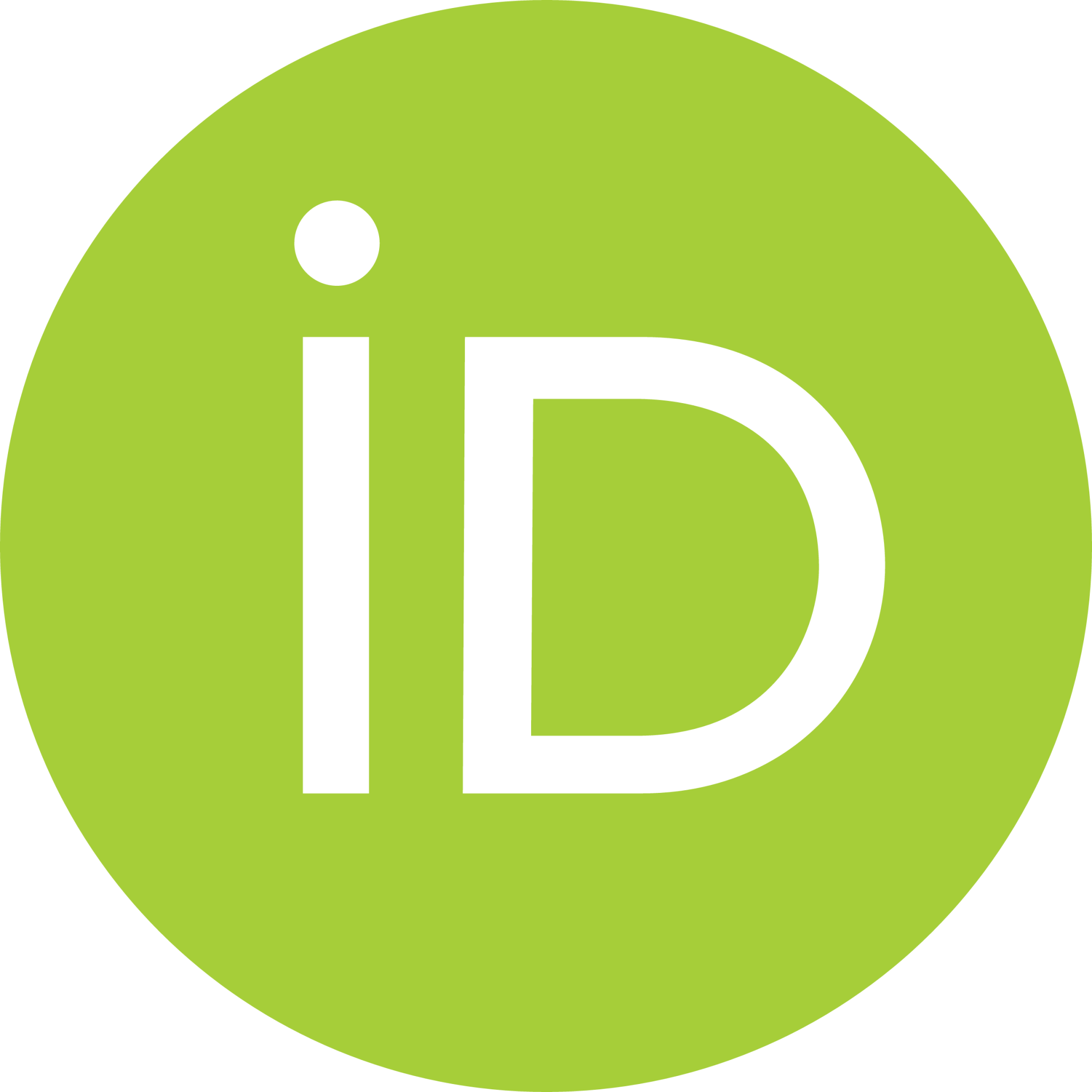}}}

\begin{document}

\preprint{APS/123-QED}

\title{Graphene is neither Relativistic nor Non-Relativistic case: Thermodynamics Aspects}

 \author{Thandar Zaw Win\orcid{0009-0006-8033-3025}, Cho Win Aung\orcid{0000-0001-5684-2854}, Gaurav Khandal, Sabyasachi Ghosh\orcid{0000-0003-1212-824X}}
\affiliation
{Indian Institute of Technology Bhilai, GEC Campus, Sejbahar, Raipur 492015, Chhattisgarh, India}
%





\begin{abstract}
    Discovery of electron hydrodynamics in graphene system has opened a new scope of theoretical research in condensed matter physics, which was traditionally well cultivated in science and engineering as a non-relativistic hydrodynamics and in high energy nuclear and astro physics as relativistic hydrodynamics. Electrons in graphene follow neither non-relativistic nor relativistic hydrodynamics. Similar to other hydrodynamical descriptions, the energy-momentum tensor of graphene also has an ideal component and a dissipating component, but in an unconventional way, so popularly, it is sometimes called as unconventional hydrodynamics. The unconventional part of the dissipating component for energy-momentum tensor is recently addressed in Phys. Rev. B 108, 235172 (2023) but its ideal component, which is connected with the thermodynamics of graphene, has not been zoomed in a very systematic way. Present article has gone through systematic microscopic calculations of thermodynamical quantities like pressure, energy density, etc., of electron-fluid in graphene and compared with corresponding estimations for non-relativistic and ultra-relativistic cases.  We have sketched the temperature and Fermi energy dependency of electron thermodynamics for graphene and other cases where the transition from Fermi liquid to Dirac fluid domain is explored. An equivalent transition for quark matter is also discussed. Interestingly, an enhancement of specific heat within the low temperature and Fermi energy region is found, which 
    may be connected to the recently observed Wiedemann-Franz Law violation.
\end{abstract}

\maketitle

\section{Introduction} 
Fluid dynamics is a macroscopic theory that describes fluid motions in terms of space-time-dependent different 
densities like energy, momentum, number, etc. 
and widely well studied in the branch of applied mathematics, science, and engineering~\cite{FD1}. When one enters into its microscopic physics, a non-relativistic kinetic theory appears for the matter of discussion~\cite{Tong1,Tong2}.
A similar kind of macroscopic and microscopic description of relativistic hydrodynamics~\cite{RHD,Astro_Hydro} is also well developed and cultivated in high energy nuclear~\cite{RHD} and astro~\cite{Astro_Hydro} physics. So, in the context of hydrodynamics, reader can find two types of frameworks -
non-relativistic hydrodynamics (NRHD), well applied in the branches of applied mathematics, science and engineering~\cite{FD1},
and relativistic hydrodynamics (RHD), well cultivated in broad band of physics from High Energy Nuclear Physics~\cite{RHD} to 
Astro Physics~\cite{Astro_Hydro} to String theory~\cite{KSS_Hydro}.
Interestingly, the hydrodynamics aspects never came into the picture in condensed matter physics or solid-state physics before the discovery of
electron hydrodynamics (eHD) in graphene system.
The experimental observations of eHD was recently pointed by Refs.~\cite{1Ku:2019lgj,2varnavides2020electron,3PhysRevB.103.125106,4PhysRevB.103.235152,5PhysRevB.103.155128,6PhysRevResearch.3.013290,7DiSante:2019zrd,8PhysRevB.103.115402,9sulpizio2019visualizing,10gallagher2019quantum,11doi:10.1126/science.aau0685,12ella2019simultaneous,13bandurin2018fluidity,18doi:10.1126/science.aad0201,14PhysRevB.98.241304,15jaoui2018departure,16doi:10.1126/science.aac8385,16doi:10.1126/science.aac8385}~. 
See Refs.~\cite{eHD1,eHD2,eHD3} for recent reviews, where the introduction of hydrodynamics in condensed matter physics has been discussed well. Refs.\cite{1Ku:2019lgj,2varnavides2020electron,3PhysRevB.103.125106,4PhysRevB.103.235152,5PhysRevB.103.155128,6PhysRevResearch.3.013290,7DiSante:2019zrd,8PhysRevB.103.115402,9sulpizio2019visualizing,10gallagher2019quantum,11doi:10.1126/science.aau0685,12ella2019simultaneous,13bandurin2018fluidity,18doi:10.1126/science.aad0201} have shown that the graphene is the best known such material, where electron hydrodynamics can be observed.
Due to the proportional relation between energy and momentum, electron motion in graphene will
not be Galilean-invariant. On the other hand, the relativistic effect of electrons cannot be expected because
their velocity is not very close to the speed of light. Hence, we can’t claim the Lorentz-invariant property of electron
motion. It opens “unconventional” hydrodynamics as neither non-relativistic nor relativistic hydrodynamics can
be applicable to this system. 

Apart from the recently discovered hydrodynamic aspect of graphene, there are some other famous properties of it, for which
it constantly attracted the attention of the scientific community~\cite{G_rev} for a few decades from when it was discovered. 
It is basically a two-dimensional system of carbon atoms with some interesting properties~\cite{G_rev} like 
\begin{itemize}
    \item zero band gap between the valance and conduction bands at Dirac points,

    \item (effectively) zero mass behavior with proportional relation between electron's energy and momentum.
\end{itemize}
From the experimental point of view, this coincidence of gap-less and mass-less nature with the hydrodynamic aspects is an observed fact, but it was a theoretical challenge to find their interconnection mechanism. According to the latest understanding~\cite{eHD1,eHD2,eHD3}, we may connect them by following links. Free electron theory of electron as Fermi gas is well applicable for most of the metals as they always have finite Fermi energy or chemical potential 
$\mu$ within the range $\mu=2-10$ eV. In the present article, we will assume the equivalence role of Fermi energy and chemical potential, which are also considered a temperature-independent parameter, but the real system may slightly differ from these assumptions. Consequently, only electrons on the Fermi-energy sphere will participate in the kinematics, thermodynamics, and transport phenomena. Other electrons, occupied below the Fermi energy sphere, will remain immobile due to the Pauli exclusion principle. So, (nearly) free electrons on the Fermi energy sphere will face obstacles of those immobile electrons as well as ions in real coordinate space, for which those free electrons follow zig-zag Ohmic type motions. In other words, electron-electron scattering will be suppressed by other dominant scatterings like electron-ions and others, which can not be removed in this Ohmic type motion of Fermionic systems having large $\mu$. These kinds of transportation provide internal energy density $\ep \propto T^2$ and specific heat $C_v\propto T$ based on the Sommerfeld expansion calculation, and also Wiedemann-Franz law is well obeyed. On the other hand, the picture may be changed for the graphene case, where $\mu$ can be decreased from the range $\mu=2-10$ eV to smaller values (even towards $\mu\rightarrow 0$ limits) by changing the doping level~\cite{eHD1,eHD2,eHD3}. So when one goes toward $\mu/T\ll 1$ domain, called Dirac fluid (DF) or Dirac liquid (DL) domain \cite{eHD1}, electron-electron scattering becomes dominating over other possible scatterings, and electron hydrodynamics or Poiseuille's motion is observed instead of Ohmic type motions. These kinds of fluid-based transportation may provide different expressions of internal energy density and may violate Wiedemann-Franz law~\cite{Sci_16_Dirac}. Present article is aimed to explore the differences in thermodynamical properties for graphene case between $\mu/T\ll 1$ and  $\mu/T\gg 1$ domains as well as compared with other cases like non-relativistic (NR) and ultra-relativistic (UR).      

If we find the earlier references, then we will get Refs.~\cite{G_thermo_1,G_thermo_2,G_thermo_3}, where 
graphene thermodynamics in the presence of magnetic field are studied. The correction term with non-interacting thermodynamics results
is addressed in Refs~\cite{G_thermo_4,G_thermo_5}. In this regard, the present work is only focused on non-interacting thermodynamics in absence of any external magnetic field. So, a straightforward statistical mechanical calculation will provide the information on graphene thermodynamics, which is basically addressed here, but we have stressed the comparison of thermodynamics for the graphene case to the other cases like NR and UR cases. Also, the transition of thermodynamics for those cases from 3 dimensions (3D) to 2 dimensions (2D) is documented. We have expressed our thermodynamical quantities like pressure ($P$), energy density ($\ep$), specific heat ($C_v$), number density ($n$) as functions of $T$ and $\mu$ in terms of Fermi integral function
\be 
f_\nu (A)=\frac{1}{\Gamma (\nu)}\int_0^\infty \frac{x^{\nu-1}}{A^{-1} e^x+1} dx ~,
\label{FIF}
\ee 
where $A=e^{\mu/k_BT}$. After getting the general expressions of thermodynamics for all cases, we have explored their $\mu/T\gg 1$ limits by using Sommerfeld expansion and $\mu/T\ll 1$ or $\mu\rightarrow 0$ limits in terms of Riemann zeta function
\be 
\zeta_\nu =\frac{1}{\Gamma (\nu)}\int_0^\infty \frac{x^{\nu-1}}{e^x-1} dx ~.
\label{RZF}
\ee 
Through these comparative studies on thermodynamics, we want to zoom in on the fact that electrons of graphene system behave as neither non-relativistic nor relativistic
cases, which is well known but sometimes misguided by the terminology - relativistic description of graphene. The present work demonstrates analytically as well as numerically the differences of graphene (G) case from NR and UR cases via different thermodynamical quantities. In parallel, it also shows that G and UR cases are hard to differentiate for some particular thermodynamical relations, connected with equation of state and equipartition law. It means that those thermodynamical outcomes may misguide us to think that electrons in graphene follow a relativistic description. This agenda is projected in this article with  the organization - first formalism part in Sec.~(\ref{sec:Form}), then numerical results part in Sec.~(\ref{sec:Res}) and at the end, summary part in Sec.~(\ref{sec:Sum}). 
%

%

%

\section{Formalism}
\label{sec:Form}
For any fluid description, we start with the energy-momentum tensor
\bea
T^{\mu\nu} &=& T_0^{\mu\nu} + T_D^{\mu\nu}~, 
\eea
where $T_0^{\mu\nu}$ is ideal part and $T_D^{\mu\nu}$ is dissipative part. 
We will see that the energy momentum tensor of graphene will be unconventional with respect
to existing NRHD~\cite{FD1} and RHD~\cite{RHD} descriptions, so we may call this new kind of
unconventional hydrodynamics as graphene hydrodynamics (GHD).
Our earlier work~\cite{Cho}, the unconventional part of dissipating component for energy momentum tensor is already exposed 
but its ideal component, which are connected with basic thermodynamics of graphene, has not been zoomed in very systematic way.
Present study is aimed to fulfill this gap. 

In terms of the building blocks - energy density $\ep$, pressure $P$, number density $n$, fluid four-velocity $u^\mu$ and metric tensor $g^{\mu \nu}$, we can write the ideal energy-momentum tensor and electron number flow as
\bea
T_0^{\mu\nu} &=&  \ep \frac{u^\mu u^\nu}{v_g^2} - P\bigg(g^{\mu\nu} -\frac{u^\mu u^\nu}{v_g^2}\bigg)~,
\nn\\
N_0^\mu &=& n \frac{u^\mu}{v_g}~.
\eea 
Here, one can design the four-velocity as $u^{\mu}=\gamma_g (v_g, \vec{u})$~ for graphene hydrodynamic (GHD) by mapping the four-velocity structure $u^{\mu}=\gamma (c, \vec{u})$ for relativistic hydrodynamics (RHD), where speed of light $c$ in RHD is basically replaced by electron Fermi velocity $v_g$ in GHD~\cite{eHD2,eHD3}. Corresponding Lorentz factor $\gamma= 1/\sqrt{1-u^2/c^2}$ in RHD will also be changed to a modified factor $\gamma_g= 1/\sqrt{1-u^2/v_g^2}$ in GHD. In static limit four-velocity, $u^{\mu}=\gamma_g (v_g, \vec{u})$ tends to $u^{\mu}=\gamma_g (v_g, \vec{0})$ and $\gamma_g= 1/\sqrt{1-u^2/v_g^2}$ tends to $1$. In this static limit, we will get $T^{00}_0=\ep$ and $T^{11}_0=T^{22}_0=T^{33}_0=P$, which is basically presenting the Pascal's law. All static limit quantities of $T^{\mu\nu}$ and $N^\mu$ are basically thermodynamical quantities like $\ep$, $P$, and $n$, which can be derived from the statistical mechanical description. For canonical ensemble or grand canonical ensemble, those macroscopic quantities can be derived from the single microscopic quantity-partition function. Alternatively, one can define them via microscopic kinetic theory-based expression of the energy-momentum tensor and electron current
\begin{equation}
T_0^{\mu\nu} =  N_s \int \frac{d^3\vec{p}}{(2\pi)^3}p^\mu v^\nu f_0~,
\label{rohan}
\end{equation}
and
\begin{equation}
    N_0^{\mu} = N_s\int \frac{d^3\vec{p}}{(2\pi)^3} v^\mu f_0~,
\label{N_mic}
\end{equation}
where $N_s=2$ is spin degeneracy factor of electron and $f_0$ is its Fermi-Dirac distribution function $f_0=1/\{\exp{(\beta(E-\mu))}+1\}$. Where $E$ = energy of fermions, $\mu$ = fermi energy i.e., the energy level up to the fermions fill or the maximum kinetic energy of fermions at absolute temperature.
The thermodynamics parameter $\beta$, which is $\beta = \frac{1}{k_B T}$ and $k_B$ is Boltzmann constant and $A$ is the fugacity of the system and which is, $A = \exp{\left({\frac{\mu}{k_BT}}\right)}$.
%

%

Using the Eqs.~(\ref{rohan}) and (\ref{N_mic}), we will calculate the basic thermodynamical quantities - energy density ($\ep$), pressure ($P$) and number density ($n$)
for different cases like G, NR, and UR, following energy and momentum relations or dispersion relations 
\bea
    E &=& pv_g~,{\rm for~G~case},
    \nn\\
    &=& p^2/(2m)~,{\rm for~NR~case},
    \nn\\
    &=& pc~,{\rm for~UR~case},
    \label{nanu}
\eea 
respectively. For 3D case, we will consider $p^2=p^2_x+p^2_y+p^2_z$ and $\int d^3p=\int 4\pi p^2 dp$, while for the 2D case, they will be considered as
 $p^2=p^2_x+p^2_y$ and $\int d^2p=\int 2\pi p dp$. We will elaborately address the 3D calculation for G case only, which will be followed for other cases - 3D NR, 3D UR, 2D G, 2D NR and 2D UR. So only final expressions of thermodynamical quantities for other cases will be written. They are addressed in the next two subsections for 3D and 2D systems, respectively. 
\subsection{Thermodynamic Quantities in 3D Space}

From the Eq. (\ref{rohan}), the energy density for the graphene system is
\begin{equation}
\epsilon_G^{3D} = T_0^{0 0} =  N_s \int \frac{d^3 p}{\left(2\pi\right)^3}\left(E \right) f_0.
\end{equation}
After using the graphene dispersion relation $E=pv_g$, we get
\bea
    \epsilon_G^{3D} &=& \frac{N_s}{2 \pi^2 v_g^3} \int_{0}^{\infty}  \frac{E^3}{A^{-1}e^{\beta E} + 1} dE~,
    \nn\\
    &=& \frac{3 N_s}{\pi^2 v_g^3} f_4\left(A\right) T^4~.
    \label{mataji}
\eea
Here, $f_4$ is Fermi integral function for $\nu=4$, given in Eq.~(\ref{FIF}).  
From the Eq.~(\ref{mataji}), the total internal energy for $V$ volume system can be written as 
\begin{equation}
    U_G^{3D} = \frac{3 N_s V}{\pi^2 v_g^3} f_4\left(A\right) T^4~.
   \label{prem}
\end{equation}
%
%
The total number of electrons in terms of the density of states can be written as
\begin{equation}
    N = \int_0^{\infty}D\left(E\right)dE f_0~.
    \label{3dnd}
\end{equation}
Here $D\left(E\right)dE$ is the  number of energy states in energy range $E$ to $E+dE$ and given in the appendix(\ref{Appendix_A}). Using it in the above equation (\ref{3dnd}), we get
\begin{align*}
    &N = N_s \frac{4\pi V}{\left(2\pi\right)^3 v_g^3} \int_{0}^{\infty}  \frac{E^2}{A^{-1}e^{\beta E} + 1} dE~.\\
\end{align*}
After converting this integral into the Fermi integral function see in the appendix(\ref{Appendix_B}), we get the final expression of number density:
\begin{equation}
    n_G^{3D} = \frac{N}{V} = \frac{N_s}{\pi^2 v_g^3} f_3\left(A\right) T^3~.
    \label{numer}
\end{equation}
One can get same expression of $n_G^{3D}$ by starting from Eq.~(\ref{N_mic}).

From the Eqs. (\ref{mataji}) and (\ref{numer}), we can write a relation between total internal energy $U=\ep V$ and total number $N=n V$ at finite volume $V$ as
\begin{equation}
    U_G^{3D} =3NT \frac{f_4\left(A\right)}{f_3\left(A\right)}.
    \label{ninad1}
\end{equation}
The electronic-specific heat for this system can be calculated by taking the derivative of total internal energy ($U$) with respect to temperature ($T$)  at constant Fermi energy ($\mu$) and constant $V$ i.e.
\begin{equation*}
    C_v = \frac{\partial U}{\partial T}\Bigg|_{\mu, V}.
\end{equation*}
In appendix~(\ref{Appendix_C}), the specific heat calculation for 2D system is addressed. Doing similar calculation for 3D case, we get
\begin{equation}
     \prescript{3D}{G}{[C_v]_{e}}  = \frac{c_v}{N} = 3 k_B \Bigg[4 \frac{f_4\left(A\right)}{f_3\left(A\right)} - \frac{\mu}{k_BT} \Bigg].
    \label{sawanra1}
\end{equation}
Similarly, from the equation (\ref{rohan}), one can calculate the pressure of 3D-G system 
\bea
    P_G^{3D} &=& T_0^{11} = N_s \int  \frac{d^3 p}{\left(2\pi\right)^3}\left(\frac{E}{3} \right) f_0~,
    \nn\\
     &=& \frac{N_s}{\pi^2 v_g^3} f_4\left(A\right) T^4.
    \label{Pg3D}
\eea
Now, after doing the same type of calculation, if the Fermi velocity of electrons in graphene $v_g$ is replaced by the speed of light $c$, then the system becomes ultra-relativistic (UR). While $[C_v]_e$ is independent of velocity, the specific heat will be the same for G and UR cases.

Similar to the calculations for 3D-G systems, if we go for the 3D-NR case calculations, then we can find 
the expressions of energy density, electronic specific heat, number density, and pressure as
\begin{align}
     \label{enr3D}
    & \epsilon_{NR}^{3D} = \frac{3 N_s}{2} \left(\frac{m}{2 \pi}\right)^{\frac{3}{2}}  f_{\frac{5}{2}}\left(A\right)  T^{\frac{5}{2}}~,\\
    \label{cv2nr3D}
    & \prescript{3D}{NR}{[C_v]_{e}} = \frac{c_v}{N} = \frac{3}{2}  k_B \Bigg[\frac{5}{2} \frac{f_{\frac{5}{2}}\left(A\right)}{f_{\frac{3}{2}}\left(A\right)} - \frac{\mu}{k_B T} \Bigg]~,\\
     \label{nnr3D}
    & n_{NR}^{3D} = N_s \left(\frac{m}{2\pi}\right)^{\frac{3}{2}} f_{\frac{3}{2}}\left(A\right) T^{\frac{3}{2}}~,\\
    \label{Pnr3D}
   & P_{NR}^{3D} = N_s\left(\frac{m}{2\pi}\right)^{\frac{3}{2}} f_{\frac{5}{2}}\left(A\right)    T^{\frac{5}{2}}.
\end{align}    

\subsection{Thermodynamic Quantities in 2D Space}
Next, we go from 3D to 2D case. So, $\frac{d^3 p}{\left(2\pi\right)^3}$ changes into $\frac{d^2 p}{\left(2\pi\right)^2}$ in the calculations.
Then, using the 2D-G dispersion relation, one can find the expressions of energy density, electronic specific heat, number density and pressure as
\begin{align}
     \label{eg2D}
    & \epsilon_G^{2D} = \frac{U}{V} = \frac{N_s}{\pi v_g^2} f_3\left(A\right) T^3~,\\
    \label{cv2g2D}
    &\prescript{2D}{G}{[C_v]_{e}} = \frac{c_v}{N} =  2k_B \Bigg[3 \frac{f_3\left(A\right)}{f_2\left(A\right)} - \frac{\mu}{k_BT} \Bigg]~,\\
    \label{ng2D}
    & n_G^{2D} = \frac{N}{V} = \frac{N_s}{2 \pi v_g^2} f_2\left(A\right) T^2~,\\
    \label{Pg2D}
    & P_G^{2D} =  \frac{N_s}{2 \pi v_g^2} f_3\left(A\right) T^3~.
\end{align}
Corresponding expressions for 2D-UR case can easily be obtained by replacing $v_g$ by $c$.

Next, using the 2D-NR dispersion relation, one can find the expressions of energy density, electronic specific heat, number density and pressure as
\begin{align}
    \label{enr2D}
    & \epsilon_{NR}^{2D} =  \frac{N_s}{2} \, \left(\frac{m}{\pi}\right) f_2\left(A\right)  T^2~,\\
    \label{cv2nr2D}
    &\prescript{2D}{NR}{[C_v]_{e}} = \frac{c_v}{N} =  k_B \Bigg[2 \frac{f_2\left(A\right)}{f_1\left(A\right)} - \frac{\mu}{k_BT} \Bigg]~,\\
    \label{nnr2D}
    &n_{NR}^{2D} = \frac{N_s}{2} \, \left(\frac{m}{\pi}\right) f_1\left(A\right)  T~,\\
    \label{Pnr2D}
    & P_{NR}^{2D} = \frac{N_s}{2} \, \left(\frac{m}{\pi}\right) f_2\left(A\right)  T^2~.
\end{align}

\section{Results}
\label{sec:Res}
%
\begin{table*}  
\caption{3D systems at $\mu\rightarrow 0$ limits: Thermodynamical quantities - $\ep$, $P$, $n$ for NR, G and UR cases.}
\begin{ruledtabular}
\begin{tabular}{ cccc } 
   & NR & G & UR\\ 
  \hline
 $\epsilon$ & $1.94 \, \left(\frac{m}{2 \, \pi}\right)^{3/2} \, \zeta_{5/2} \, T^{5/2} = (6.03 \times 10^7) T^{\frac{5}{2}}$ &$\frac{6}{\pi^2 \, v_g^3 } \, \frac{7}{8} \, \zeta_4 \, T^{4}= (0.2665 \times 10^7) T^4 $& $\frac{6}{\pi^2} \, \frac{7}{8} \, \zeta_4 \, T^{4} = 0.5756 \ T^4$\\
 $P$ & $1.29 \, \left(\frac{m}{2 \, \pi}\right)^{3/2} \, \zeta_{5/2} \, T^{5/2}  = (4.02 \times 10^7) T^{\frac{5}{2}}$ & $\frac{2}{\pi^2 \ v_g^3} \, \frac{7}{8} \, \zeta_4 \, T^{4}=(0.0888 \times 10^7) T^4 $&  $\frac{2}{\pi^2} \, \frac{7}{8} \, \zeta_4 \, T^{4}=0.1919 \ T^4$ \\ 
 $n$ & $0.59 \, \left(\frac{m}{2 \, \pi}\right)^{3/2} \, \zeta_{3/2} \, T^{3/2} = (3.55 \times 10^7) T^{\frac{3}{2}}$ & $\frac{2}{\pi^2 \ v_g^3} \, \frac{3}{4} \, \zeta_3 \, T^{3}=(0.0846 \times 10^7) T^3$ & $\frac{2}{\pi^2} \, \frac{3}{4} \, \zeta_3 \, T^{3}=0.1827 \ T^3$\\ 
\end{tabular}
\end{ruledtabular}
\label{Tab:mu03D}
\end{table*}

\begin{table*} 
\caption{3D systems at $T\rightarrow 0$ limits: Thermodynamical quantities - $\ep$, $P$, $n$ for NR, G and UR cases.}
\begin{ruledtabular}
\begin{tabular}{ cccc } 
   & NR & G & UR\\ 
  \hline 
 $\epsilon$ & $\frac{2 \ \sqrt{2} }{5 \, \pi^2 } \,  m^{3/2}  \ \mu^{5/2}=(2.09 \times 10^7) \mu^{\frac{5}{2}}$ & $\frac{1}{4 \pi^2 v_g^{3}}\, \mu^4=(0.0117 \times 10^7)\mu^4$ & $\frac{1}{4 \pi^2} \mu^4 = 0.025 \ \mu^4$\\
 $P$ & $\frac{4 \ \sqrt{2} }{15 \, \pi^2 } \,  m^{3/2}  \ \mu^{5/2}=(1.39 \times 10^7) \mu^{\frac{5}{2}}$&  $\frac{1}{12 \pi^2 v_g^3}\, \mu^4=(0.0039 \times 10^7) \mu^4$ & $\frac{1}{12 \pi^2}\, \mu^4=0.008 \ \mu^4$ \\ 
 $n$ & $\frac{2 \ \sqrt{2}}{3 \pi^2} \, m^{3/2} \ \mu^{3/2}=(3.49 \times 10^7) \mu^{\frac{3}{2}}$ & $\frac{1}{3 \pi^2 v_g^3}\, \mu^3 = (0.0156 \times 10^7) \mu^3$ & $\frac{1}{3 \pi^2}\, \mu^3=0.034 \ \mu^3$\\ 
\end{tabular}
\end{ruledtabular}
\label{Tab:Tzero3D}
\end{table*}
%

In the formalism part, we have obtained the general expressions of different thermodynamical quantities like energy density,
pressure, number density, specific heat, etc., for electron fluid inside graphene, where fluid particle - electrons follow
the dispersion relation $E=pv_g$. In parallel, we also have addressed the corresponding expressions for electrons following
NR dispersion relation $E=p^2/(2m)$ and UR dispersion relation $E=pc$ for numerical comparison, which will be presented here
in this result section. Through this comparative analysis, our aim is to highlight that graphene thermodynamics is neither
non-relativistic nor ultra-relativistic thermodynamics. Just like they are different in microscopic dispersion relations,
similarly, they are different in macroscopic thermodynamics also.

Before drawing the general expressions of three cases - G, NR, and UR for 2D and 3D systems, which are expressed in terms of Fermi integral form in the formalism part, let us first provide a tabulated description of their simple expressions in two extreme limits - (1) $\mu\rightarrow 0$  and (2) $T\rightarrow 0$. Tables (\ref{Tab:mu03D}) and (\ref{Tab:Tzero3D}) cover thermodynamics in  $\mu\rightarrow 0$ and $T\rightarrow 0$ limits respectively for 3D systems. Next, tables (\ref{Tab:mu02D}) and (\ref{Tab:T02D}) cover the same for 2D systems.

\begin{figure}
   	\centering 
    	\includegraphics[scale=0.23]{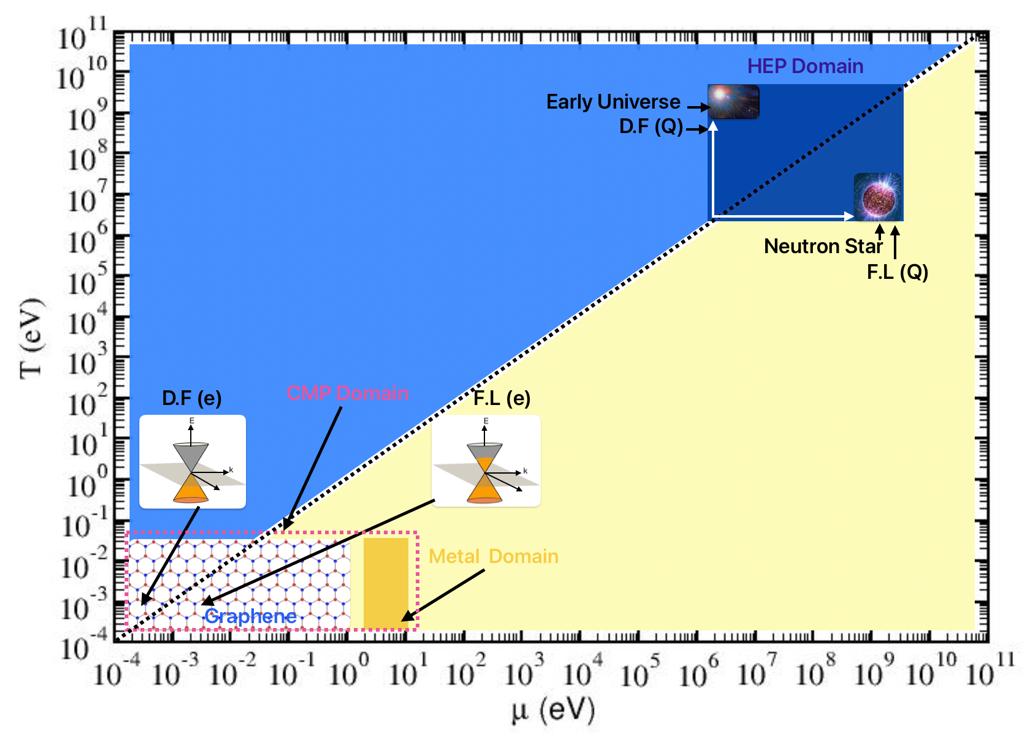}
    	\caption{The Condensed Matter Physics (CMP) and High Energy Physics (HEP) locations and their Dirac Fluid (DF) and Fermi Liquid (FL) domains in $T$-$\mu$ plane.} 
   		\label{fig:mu_T}
\end{figure}

Let us see the table (\ref{Tab:mu03D}), demonstrating $\mu\rightarrow 0$ limits of 3D system thermodynamics for NR, G, and UR cases. The reader may find the beauty of two limits - the power of T-dependence for thermodynamical quantities in $\mu\rightarrow 0$ limit remains the same in $T\rightarrow 0$ limit, the only powers of $T$ are converted to the powers of $\mu$. The constants, attached with the $T$ or $\mu$-dependent terms are different. By taking an example of energy density for 3D UR case, we can see that constants are changing from $\ep/T^4=0.57$ to $\ep/\mu^4=0.02$ due to transition from Stefan Boltzmann (SB) limit ($\mu\rightarrow 0$) to degenerate gas limit ($T\rightarrow 0$). If we take other examples also, then the reader can notice a common fact for any thermodynamical quantities of any cases that the constants are transiting from lower to
higher values during the transition from degenerate gas limit to SB limit. This may be connected with an interesting fact if we interpret it as follows. Thermodynamics is basically many body outcomes and a summation of one body quantity. After summing one body's microscopic energy scale, we get thermodynamics in terms of macroscopic energy scales $\mu$ and $T$ with some powers. So we may assume that the constants of $\ep/T^4$, $P/T^4$, $n/T^3$ in SB limit and the constants of $\ep/\mu^4$, $P/\mu^4$, $n/\mu^3$ in degenerate gas limit are basically some outcomes connected with a statistical summation. Its values are growing, which may mean that its statistical summations are growing and the thermodynamic system is going towards more thermodynamics, where the local thermalization concept can be applied. In other words, the thermodynamical system approaches the hydrodynamical system. As an example, when we take nuclear or quark matter in the SB limits, expected in high energy heavy ion collisions, we get their hydrodynamical system. Whereas, a less expectation of hydrodynamics is considered in degenerate gas limit of nuclear matter like neutron star
or dense nuclear matter, which can be produced in future accelerator facilities like NICA or CBM~\cite{CBM}. Similarly, high to low doping graphene systems exhibit a transition from non-fluid to fluid nature experimentally, which is basically a transition from degenerate to SB limit of electron thermodynamics. This non-fluid nature
is by default property for metal systems, which is always situated in degenerate gas limits (it will be more clear in specific heat discussion). Although, we should take this transition from degenerate to SB limits or non-hydro to hydro domain as a very gross picture, which is sketched in Fig.~(\ref{fig:mu_T}) because only thermodynamic knowledge can not be sufficient to realize this transition. In parallel, dissipation or interaction knowledge is required for better understanding, but
they are not a matter of interest for present work.

 According to the table (\ref{Tab:Tzero3D}), we can see that G and UR follow similar $\mu$-dependence: $\ep\propto P\propto\mu^4$, $n\propto\mu^3$ but in order of magnitude, G $\gg$ UR because $v_g^3\ll c^3$ (our expressions in the table are in natural unit, where $c=1$ is considered for UR case). Unlike the $\mu$-dependence of G or UR case, NR follows $\ep\propto P\propto\mu^{5/2}$, $n\propto\mu^{3/2}$ but in order of magnitude, it is larger than G. In terms of order of magnitude, we can conclude a gross ranking: NR $\gg$ G $\gg$ UR, which will be more clear in the graphical representation later.

\begin{table*} 
\caption{2D systems at $\mu\rightarrow 0$ limits: Thermodynamical quantities - $\ep$, $P$, $n$ for NR, G and UR cases.}
\begin{ruledtabular}
\begin{tabular}{ cccc } 
   & NR & G & UR\\ 
  \hline
 $\epsilon$ &  $\frac{1}{2} \ \left(\frac{m}{\pi}\right) \, \zeta_2 \,  T^2=(1.34 \times 10^5) T^2$ & $ \frac{2}{\pi \ v_g^2} \, \frac{3}{4} \, \zeta_3 \,  T^{3}=(0.1594\times 10^5) T^3$ & $\frac{2}{\pi} \, \frac{3}{4} \, \zeta_3 \,  T^{3}=0.5739 \ T^3$\\
 $P$ & $\frac{1}{2} \ \left(\frac{m}{\pi}\right) \, \zeta_2 \,  T^2=(1.34 \times 10^5) T^2$ & $\frac{1}{\pi \ v_g^2} \, \frac{3}{4} \, \zeta_3 \,  T^{3}=(0. 7971 \times 10^4) T^3$& $\frac{1}{\pi} \, \frac{3}{4} \, \zeta_3 \,  T^{3}=0.2870 \ T^3$ \\ 
 $n$ & undefined & $\frac{1}{\pi \ v_g^2} \, \frac{1}{2} \, \zeta_2 \,  T^{2}=(0.7184 \times 10^4) T^2$  & $\frac{1}{\pi} \, \frac{1}{2} \, \zeta_2 \,  T^{2}=0.2586 \ T^2$\\ 
\end{tabular}
\end{ruledtabular}
 \label{Tab:mu02D}
\end{table*}   

\begin{table*} 
\caption{2D systems at $T\rightarrow 0$ limits: Thermodynamical quantities - $\ep$, $P$, $n$ for NR, G and UR cases.}
\begin{ruledtabular}
\begin{tabular}{ cccc } 
   & NR & G & UR\\ 
  \hline
 $\epsilon$ & $\frac{m}{2 \pi } \ \mu^2=(8.13 \times 10^4) \mu^2 $ & $\frac{1}{3 \pi \ v_g^2 } \ \mu^3=(2.9473 \times 10^3) \mu^3$ & $\frac{1}{3 \pi} \ \mu^3=0.106 \ \mu^3$\\
 $P$ & $\frac{m}{2 \pi } \ \mu^2=(8.13 \times 10^4) \mu^2$ & $\frac{1}{3 \pi  v_g^2 } \ \mu^3=(2.9473 \times 10^3) \mu^3$ & $\frac{1}{3 \pi} \ \mu^3=0.106 \ \mu^3$ \\ 
 $n$ & $\frac{m}{\pi } \ \mu=(1.63 \times 10^5) \mu$ & $\frac{1}{2 \pi \ v_g^2 } \ \mu^2=(4.4209 \times 10^3) \mu^2$  & $\frac{1}{2 \pi} \ \mu^2=0.159 \ \mu^2$\\ 
\end{tabular}
\end{ruledtabular}
\label{Tab:T02D}
\end{table*}

%
%
%
Next, in tables~(\ref{Tab:mu02D}), (\ref{Tab:T02D}), 2D thermodynamical expressions in SB limits and degenerate for NR, G, and UR cases are addressed.
We notice the transition from $\ep\propto P\propto nT\propto T^4$ to $\ep\propto P\propto nT\propto T^3$ for G/UR case and from $\ep\propto P\propto nT\propto T^{5/2}$ to $\ep\propto P\propto nT \propto T^2$ for NR case due to 3D$\rightarrow$2D transition. A similar change in the power of chemical potential is noticed for degenerate results. The except thing is that in the SB limit, the number density of 2D NR system is found to be undefined. In general, any thermodynamical quantity depends on $f_nT^n$, where $n$ (not to be confused with number density) is an index of the Fermi integral function and power of $T$. Now in the SB limit or $\mu\rightarrow 0$ limit, we get $f_nT^n\rightarrow (1-\frac{1}{2^{n-1}})\zeta_nT^n$.
Theoretically, we get $(1-\frac{1}{2^{1-1}})=0$ for $n=1$ and $\zeta_1=\infty$, which indirectly indicate about un-physical thermodynamical quantities. Interestingly, we have limitation in real world to achieve a NR system with zero Fermi energy in 2D and 3D cases. For metal, Fermi energy or chemical potential of electron locate within $\mu=2$-$10$ eV range.

Let us go to a brief discussion on Quark Gluon Plasma (QGP), which is a more realistic UR case example than a hypothetical case - UR electron fluid.  Just after a few micro-second from the big bang, a hot quark-gluon plasma (QGP) state around temperature $T=400$ MeV or $T=700$ MeV, and zero quark chemical potential ($\mu=0$) is expected in the early universe scenario. The quark average momenta become so large because of high temperature that we can ignore its mass term, which can be considered a UR case. Photon gas or black body radiation example is famous for UR case, where internal energy density or Intensity (they have connecting relation) follows $T^4$ law, popularly known as Stefan-Boltzmann (SB) law. QGP thermodynamics at high temperatures reaches that SB limits. Energy density, pressure, number density for 3D UR case,
given in Table~(\ref{Tab:mu03D}), can be used for the quark system by replacing its relevant degeneracy factors.
For two-flavor quark has a degeneracy factor of 24. On the other hand, when one goes to $T\rightarrow 0$ limit, UR case of Table~(\ref{Tab:Tzero3D}) can be applied for the thermodynamics of the degenerate quark system, may be expected in the core of neutron star.
In this degenerate gas limit, degeneracy factor 12 which has to be considered as anti-particle probability will be highly suppressed.
    
Let us try to understand the difference between the temperature range of QGP and the graphene system.
QGP temperature is a few hundred mega electron Volt (MeV) and equivalent to $10^{12}~^0$K, which is too much larger than the temperature range of the graphene system.
Typically, $1-25$ milli electron Volt (meV) or equivalently $15-300~^0$K is the graphene system temperature range.
These two domains are nicely addressed in Fig.~(\ref{fig:mu_T}), where $T$ vs. $\mu$ plots in the log scale have covered a broad band of $T$ and $\mu$ range.
The condensed matter physics (CMP) domain specifies the temperature range of approximately $1-25$ meV and the chemical potential range of approximately $0-10$ eV.

The range of chemical potential $\mu$ from $2-10$ eV is associated with the metal Fermi energy by marking it as yellow.
Unlike metals, the Fermi energy in a graphene system can be modified or tuned through various doping methods, and its $\mu/T\ll 1$ and $\mu/T\gg 1$ domains are called Dirac fluid (DF) or Dirac liquid (DL) and Fermi liquid (FL) domains, respectively, marked by arrows in Fig.~(\ref{fig:mu_T}).
We may call early universe QGP as DF domain of quark and quark matter, anticipated in the core of neutron star as FL domain of quark. 
A rectangular domain within $T=1-400$ MeV and $\mu=0-1000$ MeV is marked as high energy physics (HEP) domain for quark. The reader can easily notice the gap between CMP and HEP domains.
\begin{figure*} 
   	\centering 
    	\includegraphics[scale=0.3]{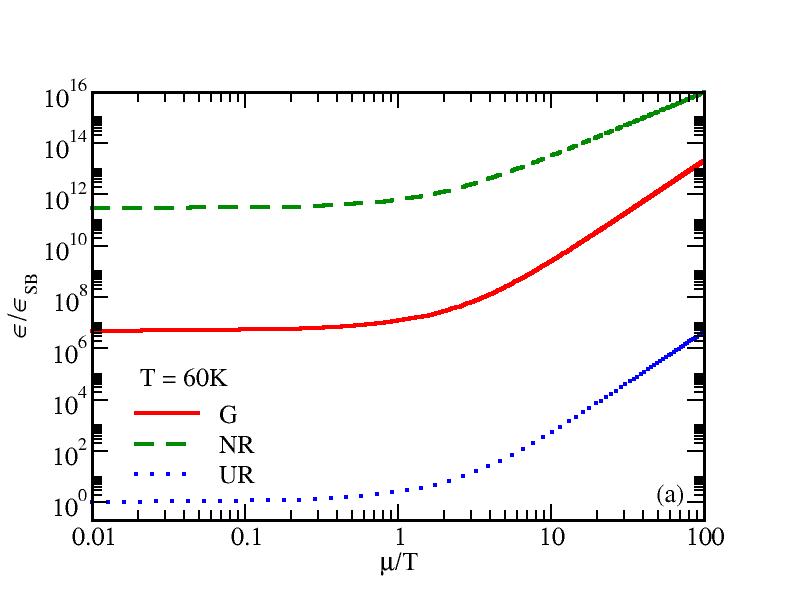}
    	\includegraphics[scale=0.3]{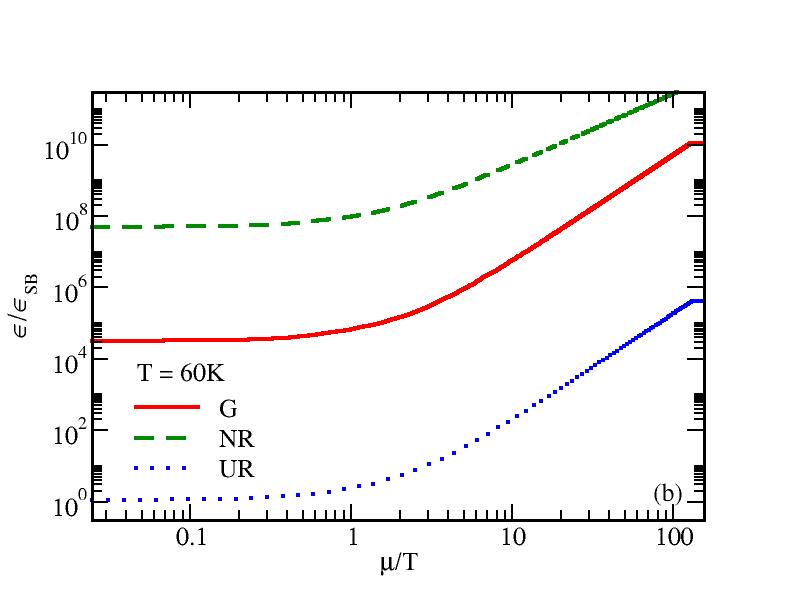}
    	\caption{The energy density in different domains to $\ep_{SB}$ with $\mu/T$ for (a) 3D and (b) 2D cases.} 
   		\label{fig:ener_den}
\end{figure*} 
\begin{figure*}   
   	\centering 
        \includegraphics[scale=0.3]{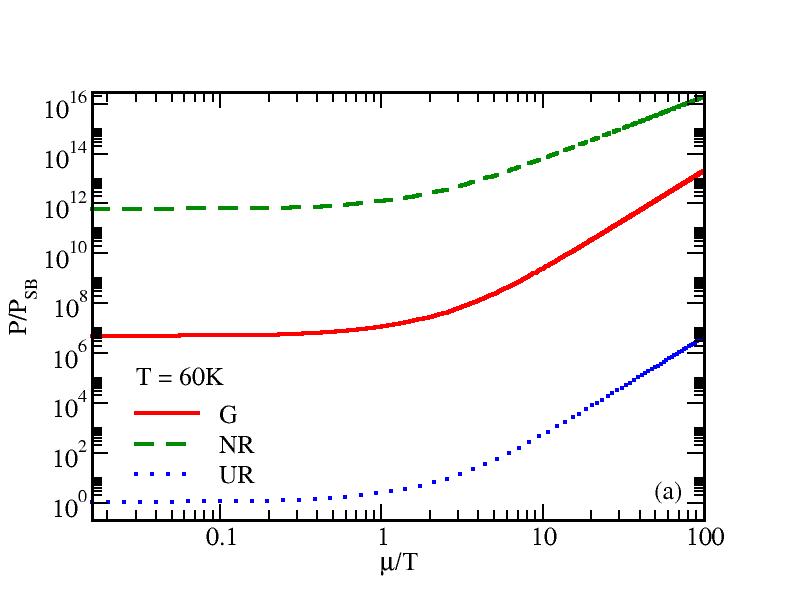}
        \includegraphics[scale=0.3]{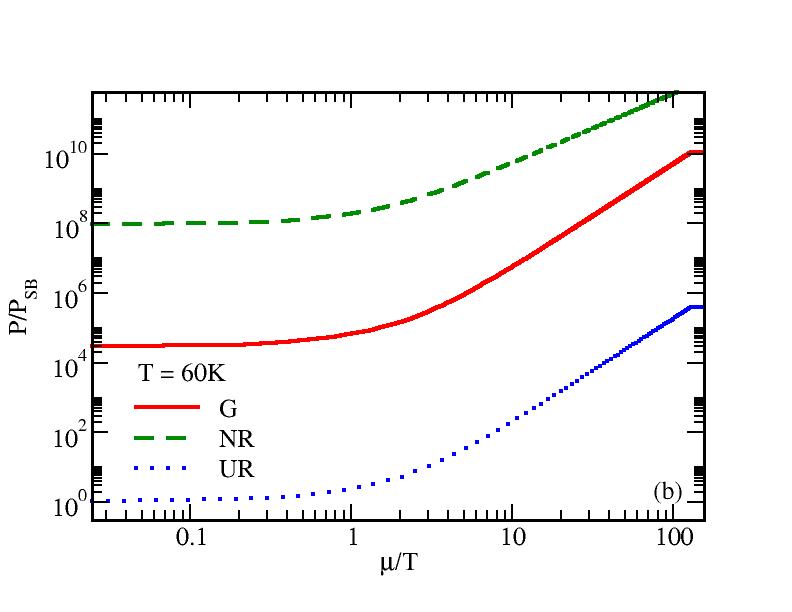}
    	\caption{The pressure in different domains to $P_{SB}$ with $\mu/T$ for (a) 3D and (b) 2D cases.} 
    	\label{fig:press}
\end{figure*} 

 %

\begin{figure*}   
   	\centering 
    	\includegraphics[scale=0.3]{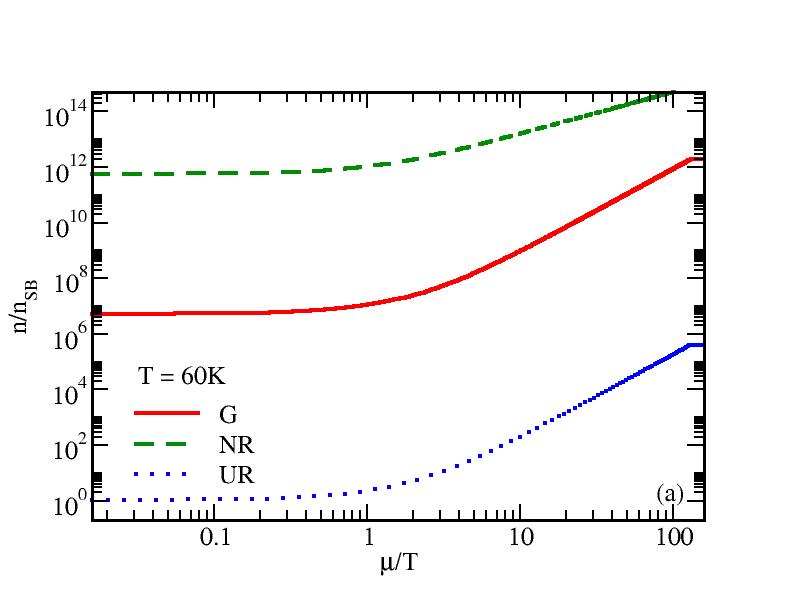}
    	\includegraphics[scale=0.3]{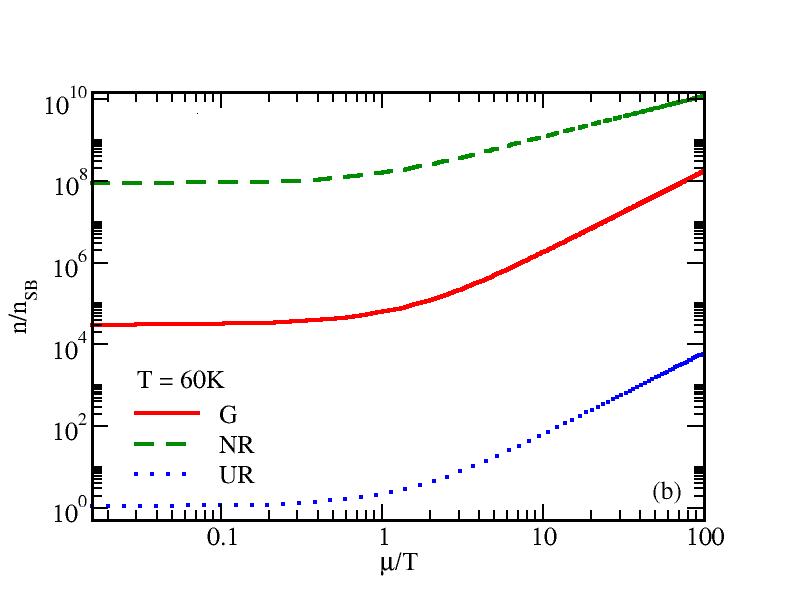}
       \caption{The number density in different domains to $n_{SB}$ with $\mu/T$ for (a) 3D and (b) 2D cases.} 
   		\label{fig:num_den}
\end{figure*}

We understand the scale gap in the $T$-$\mu$ plane between UR case of the quark system and the G case of the electron system. So, to see in the same $T$-$\mu$ scale, we must consider a hypothetical electron ultra-relativistic fluid (URF) to make an equal footing comparison with electron graphene fluid (GF). 
We can call $\ep$, $P$, $n$ of UR case in Tables~(\ref{Tab:mu03D}) and (\ref{Tab:mu02D}) as $\ep_{SB}$, $P_{SB}$, $n_{SB}$. Next putting $v_g=c=1$ in $\ep$, $P$, $n$ of Eqs.~(\ref{prem}), (\ref{numer}), (\ref{Pg3D}), we get
\bea 
   U_{UR}^{3D} &=& \frac{3 N_s V}{\pi^2 } f_4\left(A\right) T^4,
\nn\\
  n_{UR}^{3D} &=& \frac{N}{V} = \frac{N_s}{\pi^2 } f_3\left(A\right) T^3~,
\nn\\
    P_{UR}^{3D} &=& \frac{N_s}{\pi^2} f_4\left(A\right) T^4~.
    \label{PUR3D}
\eea 
Normalizing Eqs.~(\ref{PUR3D}) by their SB limits, we have sketched them by the blue dotted lines in the left panel of Figs.~(\ref{fig:ener_den}), (\ref{fig:press}), (\ref{fig:num_den}) which become one in the domain $\mu/T\ll 1$, as expected. Here, the temperature is fixed at $T=60^0$ K, and $\mu$ is varied. Interestingly, we noticed that the main $\mu/T$ dependence in thermodynamical quantities is coming beyond the $\mu/T=1$. It is Fermi integral functions, which are the main source of $\mu/T$ dependence. Next, the red solid lines in the left panel of Figs.~(\ref{fig:ener_den}), (\ref{fig:press}), (\ref{fig:num_den}) represent the graphene thermodynamics, where Fermi velocity $v_g=0.006$ is considered. We have considered in-between constant values $v_g=0.006$ from the Fermi velocity range $v_g=1-3\times 10^6$ m/s or $v_g=0.003-0.01$ (in natural unit) in graphene system~\cite{vg_Nature}. The noticeable point is that $\mu/T$ dependence of UR and G thermodynamics are the same but $G\gg UR$ due to the $1/v_g^3\approx 5\times 10^6$ term. Next, we use Eqs.~(\ref{enr3D}), (\ref{nnr3D}), (\ref{Pnr3D}) to draw NR thermodynamical quantities (green dashed lines). In order of magnitude, we get ranking UR $\ll$ G $\ll$ NR.

A similar trend can be noticed for 2D cases with similar ranking URF $\ll$ GF $\ll$ NRF. Only for the transition from 3D to 2D their orders of magnitude are shifted toward lower values.

\begin{figure*}  
   	\centering 
    	\includegraphics[scale=0.3]{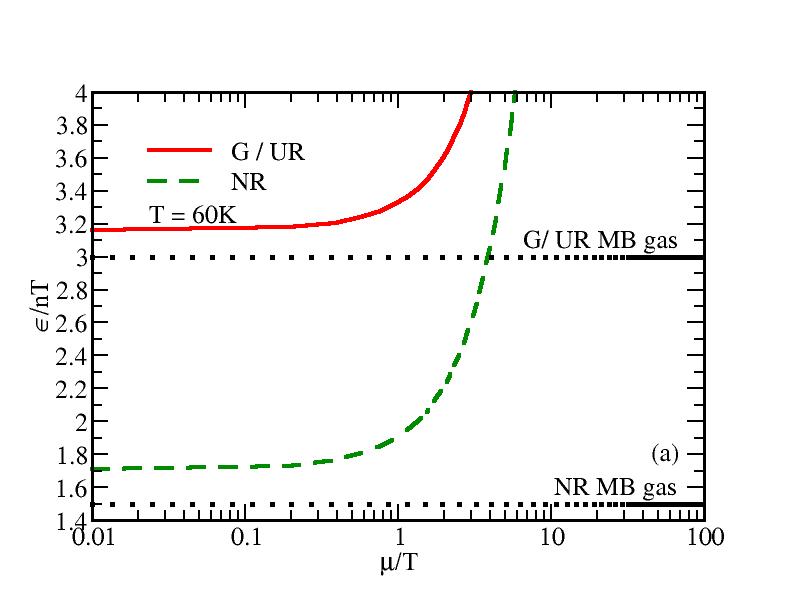}
    	\includegraphics[scale=0.3]{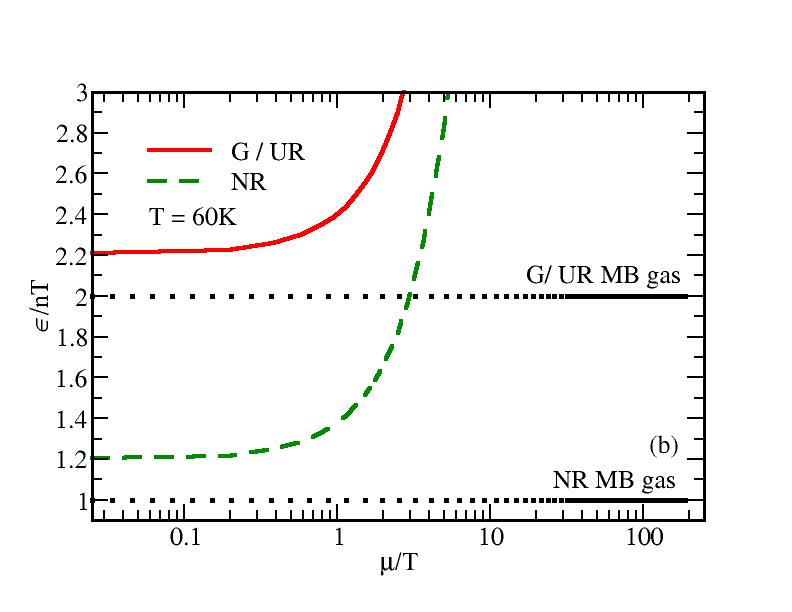}
    	\caption{The ratio of the energy density and number density for (a) 3D and (b) 2D cases.} 
   		\label{fig:ebnT}
\end{figure*}

There is a famous equi-partition law, which states that energy density by number density $\ep/n$ will be equal to the number of degrees of freedom times $\frac{1}{2}k_BT$. If we define degrees of freedom as $D$, then $\frac{\ep}{n}=\frac{D}{2}T$ (in natural unit $k_B=1$). Therefore, for 3D system having three degrees of freedom, $\frac{\ep}{nT}=\frac{3}{2}=1.5$ and for 2D system having two degrees of freedom, $\frac{\ep}{nT}=\frac{2}{2}=1$. These two horizontal dotted lines at $1.5$ and $1$ are drawn in the left and right panels of Fig.~(\ref{fig:ebnT}). This is true for ideal classical gas, whose constituent particles follow Maxwell Boltzmann (MB) statistics and NR dispersion relation, but for ideal quantum electron gas, whose constituent particles follow FD statistics, the equi-partition law becomes little modified. The $\frac{\ep}{nT}$ does not remain constant; it will change with $\mu/T$ as shown by the green dashed line for NR electron gas.
Interestingly, the major changes or diverging trend is noticed beyond $\mu/T=1$ or FL domain, while in the DF domain, it is almost
saturated towards the value $1.72$,
which is slightly greater than 1.5. This small
deviation from 1.5 to 1.72, shown in the left panel of Fig.~(\ref{fig:ebnT}), is basically due to the quantum statistical effect. Same deviation from 1 to 1.2 can be seen in the right panel of Fig.~(\ref{fig:ebnT}) for the 2D NR system. Here, although we get the value 1.2 numerically from the Fermi-integral function, we can realize the saturating value at $\mu\rightarrow 0$, but the analytical calculation of $n$ at $\mu=0$ becomes undefined due to $0\times\infty$.

\begin{figure*}  
   	\centering 
     \includegraphics[scale=0.3]{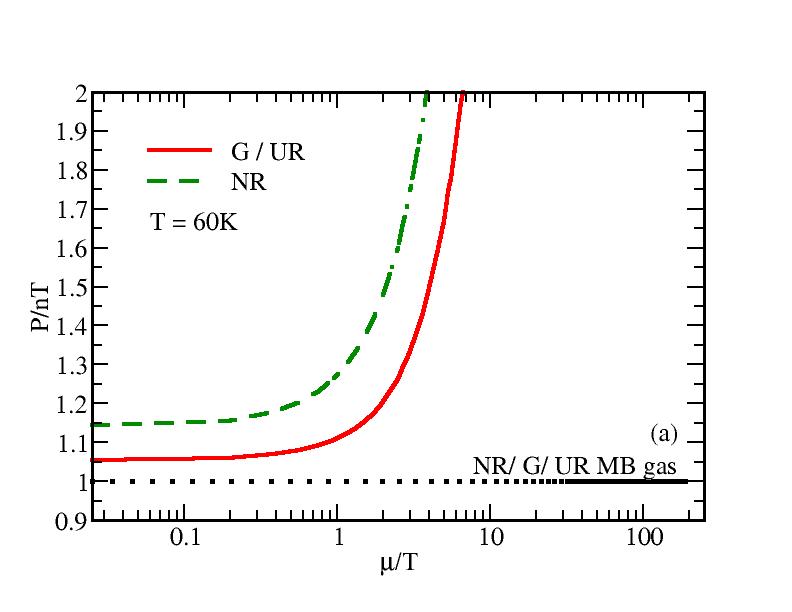}
       \includegraphics[scale=0.3]{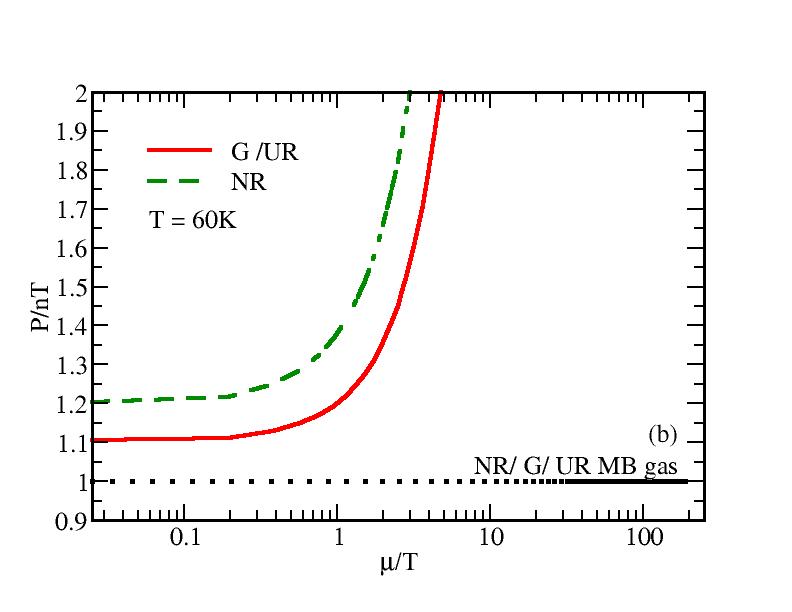}
    	\caption{The ratio of the pressure and number density for (a) 3D and (b) 2D cases.} 
    		\label{fig:pbnt} 
\end{figure*} 
Next, when we go for UR and G cases, interestingly, both follow the same equi-partition law as demonstrated by the red line
of Fig.~(\ref{fig:ebnT}). The reason for merging between UR and G in $\frac{\ep}{nT}$ is as follows. The main difference between UR and G thermodynamical quantities is due to the role of $c$ and $v_g$ in their respective expressions. During calculating ratio $\frac{\ep}{nT}$, they are canceled, so their ratio becomes exactly the same. Here also, for ideal classical gas, one can formulate equi-partition law
for UR gas as follows. Energy density by number density $\ep/n$ will be equal to the number of degrees of freedom times $T$ instead of $\frac{1}{2}T$. If we define degrees of freedom as $D$, then $\frac{\ep}{n}=DT$, which means $\frac{\ep}{nT}=3$ for 3D system and $\frac{\ep}{nT}=2$ for 2D system as drawn in left and right panels of Fig.~(\ref{fig:ebnT}). Due to quantum statistical effect, $\frac{\ep}{nT}$ UR or G gas will shift from $3$ to $3.15$ for 3D case and from $2$ to $2.21$ for the 2D case. We can define the quantum statistical factor as $Q_{G/UR}$ and $Q_{NR}$ for G/UR and NR system at $\mu=0$ to formulate equi-partition law:
\bea 
\frac{\ep}{nT}&=&D\times Q_{G/UR}~{\rm for~G/UR~case}~,
\nn\\
&=&\frac{D}{2}\times Q_{NR}~{\rm for~NR~case}~,
\eea 
with 
\bea 
Q_{G/UR}&=&\frac{(1-1/2^3)}{1-1/2^2}\frac{\zeta_4}{\zeta_3}~{\rm for~3D~ system}~,
\nn\\
&=&\frac{(1-1/2^2)}{1-1/2}\frac{\zeta_3}{\zeta_2}~{\rm for~2D~ system}~,
\nn\\
Q_{NR}&=&\frac{(1-1/2^{3/2})}{1-1/2^{1/2}}\frac{\zeta_{5/2}}{\zeta_{3/2}}~{\rm for~3D~ system}~,
\nn\\
&=&1.2~{\rm for~2D~ system}~.
\label{Q_mu0}
\eea 
These quantum statistical factors almost become independent of $\mu/T$ in DF domain, and we may safely consider the values of $\mu=0$, given in Eq.~(\ref{Q_mu0}). However, in the FL domain, one should consider quantum statistical factors in terms of Fermi integral functions as
\bea 
Q_{G/UR}&=&\frac{f_4}{f_3}~{\rm for~3D~ system}~,
\nn\\
&=&\frac{f_3}{f_2}~{\rm for~2D~ system}~,
\nn\\
Q_{NR}&=&\frac{f_{5/2}}{f_{3/2}}~{\rm for~3D~ system}~,
\nn\\
&=&\frac{f_2}{f_1}~{\rm for~2D~ system}~.
\label{Q_mu}
\eea 

After exploring the equi-partition law, let us come to the equation of state (EoS), relating $P$ and $n$. 
For ideal classical gas, we know that though $\frac{\ep}{nT}$ depends on dispersion relations $E\propto p$ or $E\propto p^2$ as well as dimension $D$, but $\frac{P}{nT}$ remains independent of those. 
Ideal gas EoS always becomes $\frac{P}{nT}=1$ for any cases 3D/2D or NR/G/UR, which are shown by dotted horizontal lines in Fig.~(\ref{fig:pbnt})(a) and (b).
Now actual 3D, 2D electron NR, G, and UR gas, following FD statistics, will have EoS $\frac{P}{nT}>1$ due to the same quantum statistical factors, described by Eq.~(\ref{Q_mu0}) for DF and Eq.~(\ref{Q_mu}) for FL.


%
\begin{figure*}   
  	\centering 
    \includegraphics[scale=0.3]{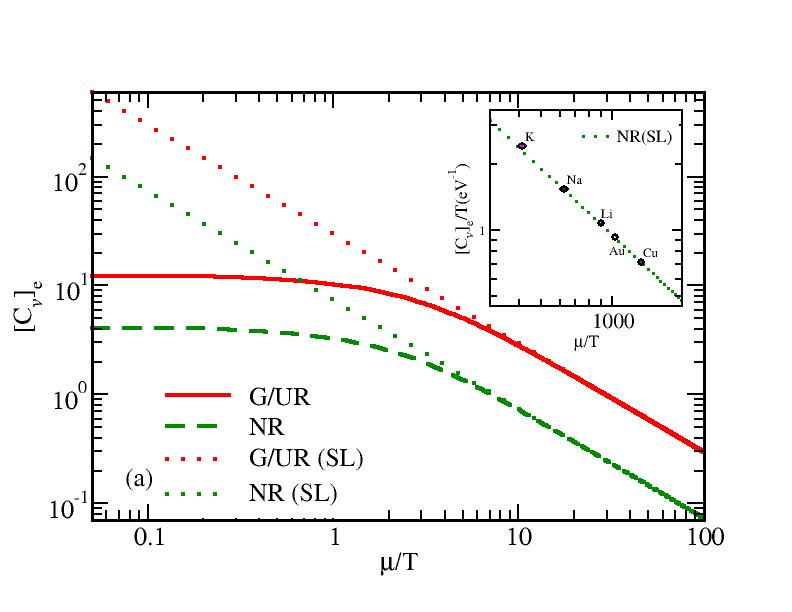}
    \includegraphics[scale=0.3]{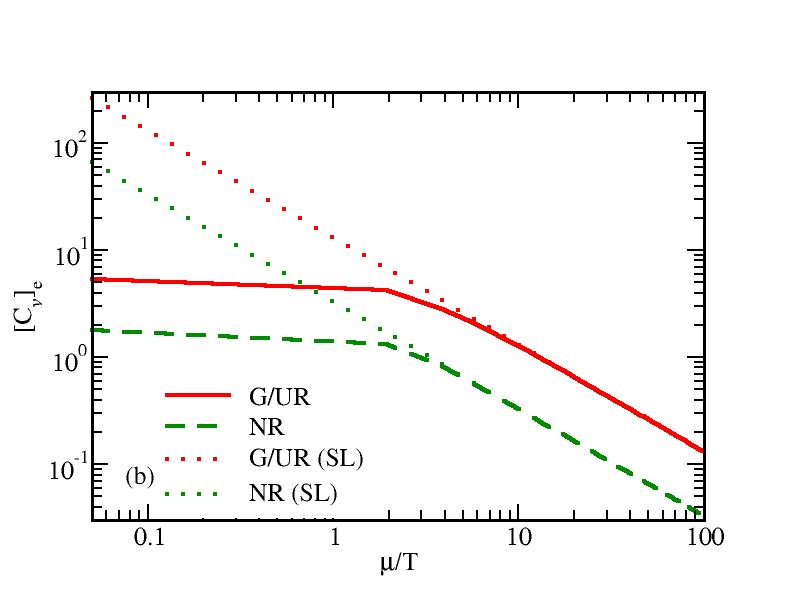}
    \caption{Specific heat of NR (green dashed line) and G (red solid line) for (a) 3D and for (b) 2D cases, and their Sommerfeld limits (SLs) (dotted lines). The inset shows the comparison of the SL results of the NR case (green dotted line) and the experimental (dotted symbols) of some metals at their respective Fermi energy~\cite{Solid_Book}.} 
 	\label{fig:CV2SLgen}
 \end{figure*} 
Finally, we constructed the directly measurable thermal quantity, the electronic specific heat capacity of G and NR in Fig.~(\ref{fig:CV2SLgen}) in terms of $\mu/T$. Here, the green lines are represented for NR, and the red lines for G or UR cases. Since $\frac{\ep}{n}=\frac{U}{N}$ for both G and UR are the same, their temperature derivative, which is basically specific heat per electron, also becomes the same.   
In $\mu/T >> 1$ or FL domain, Sommerfeld's limits (SL) can be applied to Eqs.~(\ref{sawanra1}), (\ref{cv2nr3D}), (\ref{cv2g2D}) and (\ref{cv2nr2D})
for 3D-G, 3D-NR, 2D-G, and 2D-NR systems, respectively and their corresponding expressions will be converted into very simple forms:
\bea 
\prescript{3D}{G}{[C_v]_{e}} &=& 3\pi^2 \ \frac{T}{\mu}~,
\nn\\
\prescript{3D}{NR}{[C_v]_{e}} &=& \frac{3 \pi^2}{4} \ \frac{T}{\mu}~,
\nn\\
\prescript{2D}{G}{[C_v]_{e}} &=& \frac{4 \pi^2}{3} \ \frac{T}{\mu}~,
\nn\\
\prescript{2D}{NR}{[C_v]_{e}} &=& \frac{\pi^2}{3} \ \frac{T}{\mu}~.
\label{CV_SL}
\eea 

They are plotted by dotted lines in the figure, whose trends indicate the standard $C_v\propto \frac{T}{\mu}$ relation as expected for metal in the low-temperature domain. The observation is that the specific heat of G-3D is four times larger than that of the NR-3D case. 
The proportional constant of $C_v\propto T$ will be a bit different from standard Solid state physics book~\cite{Solid_Book}, where number density and volume are kept constant, but we have used the definition, where $\mu$ and volume are kept constant. The reader may comprehend the constant number density and volume definition of specific heat as a solid state-based definition, whereas the constant $\mu$ and volume definition of specific heat may be considered a fluid-based definition. We know that (almost) in-compressible solid-based formalism for electrons in the metal, thermodynamical quantities like $P$, $V$, $N$ never be a matter of interest to calculate as they always remain constant but its $U$ and $C_v$ are normally calculated by using quantum statistical formalism in the $\mu/T\gg 1$ limit, which is basically free electron theory, practiced in solid state physics book~\cite{Solid_Book}. We can not decrease Fermi energy $\mu$ for the metal case as their approximate range is $\mu=2$-$10$ eV, which almost remains constant
with $T$. The inset in Fig.~(\ref{fig:CV2SLgen})(a) demonstrates to examine the standard validity ~\cite{Solid_Book} of the theoretical results for SL(NR) case (green dotted) with the experimental data (solid circles) for different metals (K, Na, Li, Au, and Cu). 

It was graphene, where carrier density $n$ can be tuned by changing $T$ and $\mu$. So one can access the range lower than $\mu/T=1$, and even can reach DF domain - $\mu/T\ll 1$. In this DF region, SL $C_v\propto T$ is invalid; one should use the general expression given in Eq.~(\ref{cv2g2D}). From that general expression, an elevation in specific heat is observed within the low doping region for the actual 2D-G case. Such enhancement consequently results in a magnification of thermal conductivity, as thermal conductivity is directly proportional to specific heat, which may be related to the Wiedemann-Franz law violation \cite{Sci_16_Dirac}.

\section{Summary}
\label{sec:Sum}
    In summary, the present work is focused on thermodynamic aspects of the graphene system after getting inspiration from the recent discovery of electron hydrodynamics in the graphene system. That experimental discovery has opened a new scope of analytic calculations in condensed matter physics, which was traditionally well cultivated in science and engineering as non-relativistic hydrodynamics and in high energy nuclear and astro physics as relativistic hydrodynamics. An interesting fact about electrons in graphene is that it follows neither non-relativistic nor relativistic hydrodynamics. Present work has highlighted that fact by concentrating on the ideal part of the energy-momentum tensor and current of electrons in graphene, whose static limit quantities are basically thermodynamical quantities like pressure, energy density, etc. The present article has gone through systematic microscopic calculations of that graphene thermodynamics and compared them with corresponding estimations for non-relativistic and ultra-relativistic cases. Firstly, at low temperatures (around $60^0$ K), Dirac fluid (DF) and Fermi liquid (FL) domains are respectively marked for low and high Fermi energy, and then electron thermodynamics for graphene and other cases are sketched to see their estimations in those two domain. From the final expressions and graphs of those thermodynamics for 3 dimensions (3D) and 2 dimensions (2D) graphene (G), non-relativistic (NR) and ultra-relativistic (UR) cases, we can highlight our findings in a few bullet points as:
\begin{itemize}
    \item We find the ranking UR $\ll$ G $\ll$ NR and 2D $\ll$ 3D in terms of order of magnitude of thermodynamical quantities - pressure, energy density and number density.
    \item Interestingly, G and UR both follow the same equation state $\frac{P}{nT}=Q$ and equi-partition law $\frac{\ep}{nT}=D~Q$, where $D$ is the dimension of systems and $Q$ is its corresponding quantum statistical factors in terms of Fermi integral function for FL domain and Riemann-Zeta function for DF domain. On the other hand, NR case follows equation state $\frac{P}{nT}=Q$ and equi-partition law $\frac{\ep}{nT}=\frac{D}{2}~Q$, where $Q$ is its corresponding quantum statistical factors for NR case.  
    \item  We have interpreted  the enhancement of specific heat in the low doping region for the actual 2D-G case. This enhancement leads to an increase in thermal conductivity because the thermal conductivity is being proportional to specific heat. This interpretation may be connected to the experimentally observed Wiedemann-Franz Law violation in DF domain of the graphene system.
\end{itemize}
    We have considered mili-electron Volt order temperature and electron Volt order Fermi energy for all cases, including the UR case. However, as a realistic example, quark matter at few hundred Mega electron Volt temperature and thousands Mega electron Volt Fermi energy or chemical potential is a good example. Similar to FL to DF transition in the condensed matter physics domain, an equivalent transition for quark matter in the high energy physics domain is also discussed.


\begin{acknowledgments}
This work was partly (T.Z.W. and C.W.A.) supported by the Doctoral Fellowship in India (DIA) program of the Ministry of Education, Government of India. The authors thank the other members of eHD club Sesha P. Vempati, Ashutosh Dwibedi, Narayan Prasad, Bharat Kukkar, and Subhalaxmi Nayak.
\end{acknowledgments}


%
%
\appendix

\section{Density of States}
\label{Appendix_A}
%
The density of states is nothing but the total number of energy states per unit energy interval. If the total number of energy states in energy range $E$ to $E + dE$ are $D\left(E\right)dE$ then the density of states will be 
\begin{equation}
    g\left( E \right) = \frac{D\left(E\right)dE}{dE}~.
\end{equation}

\textbf{\underline{Case:1.}} for 3D Graphene,
\begin{equation}
    D\left(E\right)dE = N_s \frac{4\pi V}{h^3 v_g^3} E^2 dE~.
\end{equation}

\textbf{\underline{Case:2.}} for 3D Non-Relativistic,
\begin{equation}
     D\left(E\right)dE = N_s 2\pi V \left(\frac{2m}{h^2}\right)^{\frac{3}{2}}\sqrt{E} \, dE~.
\end{equation}

\textbf{\underline{Case:3.}} for 2D Graphene,
\begin{equation}
    D\left(E\right)dE = N_s \frac{2\pi S}{h^2 v_g^2} E \, dE~.
\end{equation}

\textbf{\underline{Case:4.}} for 2D Non-Relativistic,

\begin{equation}
     D\left(E\right)dE = N_s\frac{2\pi S}{h^2} m\, dE~.
\end{equation}

In the above expressions, $V$ and $S$ represent the volume and area in position space, respectively.


\section{Fermi-Dirac Function}
\label{Appendix_B}
The Fermi-Dirac integral Function
\begin{align}
   f_\nu (A)=\frac{1}{\Gamma (\nu)}\int_0^\infty \frac{x^{\nu-1}}{A^{-1} e^x+1} dx~. 
   \label{mom}
\end{align}
%
\textbf{\underline{Case:1}}\\
When $\mu\rightarrow0$ (i.e., $\mu/T\ll 1$), then the Fermi-Dirac function can be written in a series form which is 
\begin{equation}
     f_\nu (A)=  \left( 1-\frac{1}{2^{n-1}} \right) \zeta_n~.
\end{equation}
%
\textbf{\underline{Case:2}}\\
When the temperature is very small, and the Fermi energy has some finite values (i.e., $\mu/T\gg 1$), then the Fermi-Dirac function can be written according to Sommerfeld's lemma, which gives the expression of the function as
\begin{widetext}
\begin{equation}
    f_\nu\left(A\right) = \frac{\alpha^\nu}{\Gamma\left(\nu + 1\right)} \Bigg[1 + \nu \left(\nu -1\right)\frac{\pi^2}{6} \frac{1}{\alpha^2}\\ 
    + \nu \left(\nu -1\right)\left(\nu -2\right)\left(\nu -3\right)\frac{7\pi^4}{360} \frac{1}{\alpha^4}+ \dots \Bigg]~,
\end{equation}
\end{widetext}

where $\alpha = \ln{A} = \frac{\mu}{k_BT}$~.
 

\begin{itemize}
   
\item {The derivative of Fermi function with respect to Fermi energy}

\begin{equation}
    \frac{\partial f_\nu \left(A\right)}{\partial \mu} = \beta  f_{\nu-1} \left(A\right)~.
\end{equation}

\item {The derivative of Fermi function with respect to temperature}

\begin{equation}
     \frac{\partial f_\nu{} \left(A\right)}{\partial T} = \frac{1}{A}  f_{\nu-1} \left(A\right) \frac{\partial A}{\partial T}~.
      \label{sd}
\end{equation}
And the one identity for the function can be written as
\begin{equation}
    \frac{1}{A} \frac{\partial A}{\partial T} = - \mu \beta^2 k_B = -\frac{\mu}{k_B T^2}~.
    \label{rama}
\end{equation}
\end{itemize}

\section{Electronic Specific Heat Capacity}
\label{Appendix_C}

{For 2D system graphene, which follows the dispersion relation in Eq. (\ref{nanu}), considering this type of system with the help of the density of state method, we can calculate the number density, total internal energy, and the electronic-specific heat.}\par

So, the number of energy states in the energy range $E$ to $E+dE$ is written as
\begin{equation}
    D\left(E\right) dE = N_s \frac{S}{\left(2\pi \right)^2} \frac{2\pi}{v_g^2} E dE~,
\end{equation}
where S is the surface area of the electron fermionic system. Now, the total number of particles at any value of temperature can be calculated as 
\begin{equation}
    N = \int_0^{\infty}D\left(E\right) dE f_0\left(E\right)~.
    \label{3DnD}
\end{equation}
After plugging the value of $ D\left(E\right) dE $ in the above equation, we get
\begin{equation}
    n = \frac{N}{S} = \frac{N_s}{2 \pi v_g^2} \int_{0}^{\infty}  \frac{E}{A^{-1}e^{\beta E} + 1} dE~.
\end{equation}
For 2D, $n$ is the number of particles per unit area, so $S$ is surface area. After solving this integration, we get the final more general expression of number density, which is given by
\begin{equation}
     N =  N_s \frac{2 \, \pi \, S}{h^2 \,  v_g^2} \frac{\Gamma \left(2\right)}{\beta^2} f_2\left(A\right)~,
     \label{gauri3}
\end{equation}
\begin{equation}
    n = N_s \frac{2  \, \pi}{ h^2  \, v_g^2} f_2\left(A\right) T^2~.
\end{equation}
Now, the total internal energy of a system can be calculated as
\begin{equation}
    U = \int_0^{\infty}  D\left(E\right) dE\,  E \, f_0\left(E\right)~. 
\end{equation}
After plugging the value of $ D\left(E\right) dE$ in the above equation and solving the integration, we get the final more general expression of total internal energy which is given by
\begin{equation}
    U = N_s \frac{2 \pi S}{h^2 v_g^2} \frac{\Gamma \left(3\right)}{\beta^3} f_3\left(A\right).
    \label{raman3}
\end{equation}
Now, from equation (\ref{gauri3}) and (\ref{raman3}), we get
 \begin{equation}
     U = 2 N T \frac{f_3\left(A\right)}{f_2\left(A\right)}.
     \label{ninad3}
 \end{equation}
Now we can calculate the specific heat capacity with the help of total internal energy  expression by taking the derivative of $U$ with respect to temperature for a constant chemical potential $\mu$ and area $S$ in 2D. So, specific heat capacity is
\begin{equation}
     C_v = \frac{\partial U}{\partial T}\Bigg|_{\mu, S}~.
\end{equation}
From the Eq.(\ref{raman3}), total internal energy is 
\begin{align}  
     \prescript{2D}{G}{U} &= N_s \frac{2 \pi S}{h^2 v_g^2} \frac{\Gamma \left(3\right)}{\beta^3} f_3\left(A\right)~,\\
     &  = N_s \frac{2 \pi S}{h^2 v_g^2} \Gamma \left(3\right)k_B^3 T^3 f_3\left(A\right)~,\\
     & = \gamma T^3  f_3\left(A\right),
\end{align}
where 
$$  \gamma =  N_s \frac{2 \pi S}{h^2 v_g^2} \Gamma \left(3\right)k_B^3 . $$
Now, after taking the derivative of just the above equation with respect to $T$ at constant Fermi energy and volume, we get
\begin{align}
      C_v &= \gamma  \Bigg[3 T^2 f_3\left(A\right) +  T^3 f_2\left(A\right) \frac{1}{A} \frac{\partial A}{\partial T} \Bigg]~,\\
      & = \gamma T^2 f_2 \left(A\right) \Bigg[3 \frac{f_3\left(A\right)}{f_2\left(A\right)} +  T \frac{1}{A} \frac{\partial A}{\partial T} \Bigg].
      \label{bolu3}
\end{align}
After using the identity (\ref{rama}), and also from the Eq.(\ref{gauri3}), the relation between $N$ and $\gamma$ can be written as
\begin{align}
     \gamma T^2 f_2 \left(A\right) = 2N k_B,
\end{align}
and now Eq.(\ref{bolu3}) becomes
\begin{align}
    C_v = 2 N k_B \Bigg[3 \frac{f_3\left(A\right)}{f_2\left(A\right)} - \frac{\mu}{k_BT} \Bigg].
\end{align}
Now, the final expression of electronic-specific heat per particle can be written in a more general form as
\begin{equation}
    \prescript{2D}{G}{[C_v]_e} = \frac{c_v}{N} =  2k_B \Bigg[3 \frac{f_3\left(A\right)}{f_2\left(A\right)} - \frac{\mu}{k_BT} \Bigg].
    \label{sawanra3}
\end{equation}

\bibliographystyle{apsrev4-2}
\bibliography{new1new}

\end{document}